\documentclass[twocolumn,superscriptaddress,aps,preprintnumbers,amsmath,amssymb,sort&compress,nofootinbib]{revtex4-2}

\usepackage{amsfonts,amsmath,amssymb,ascmac,bm,tensor}
\usepackage{fnpct} 
\usepackage{comment}
\usepackage{ifpdf}
\usepackage{slashed}
\usepackage{color}
\usepackage[mathscr]{eucal}
\usepackage[utf8]{inputenc}
\usepackage{physics}
\usepackage{cancel}

\ifpdf
  \usepackage{graphicx}     
  \usepackage[bookmarksopen,colorlinks=true,linkcolor=bblue,citecolor=bblue,urlcolor=ppink]{hyperref}
\else     
\fi

\definecolor{red}{rgb}{1,0,0}
\definecolor{darkred}{rgb}{0.6,0,0}
\definecolor{darkgreen}{rgb}{0.992447,0.623778,0.034597}
\definecolor{ppink}{rgb}{1,0.4,0.4}
\definecolor{bblue}{rgb}{0.284602,0.317763,0.963947}
\definecolor{purple}{rgb}{0.5 ,0, 0.7}

\newcommand{\ee}{\text{e}}
\newcommand{\Pl}{\text{Pl}}

\newcommand{\GW}{\text{GW}}

\allowdisplaybreaks[1]

\begin{document}


\title{
Traces of a Heavy Field in Gravitational Waves
}

\author{Keisuke Inomata}
\affiliation{Kavli Institute for Cosmological Physics, The University of Chicago, Chicago, IL 60637, USA}

\begin{abstract}
\noindent
We discuss gravitational waves (GWs) induced by a heavy spectator field that starts to oscillate during inflation. 
During the oscillation of the spectator field, its effective mass can also oscillate in some potentials.
This mass oscillation can resonantly amplify the spectator field fluctuations.
We show that these amplified fluctuations can induce large GWs, which could be investigated by future gravitational wave observations.
This kind of induced GW can be produced even if the spectator field does not have any interaction with other fields except for gravitational interaction.
\end{abstract}

\date{\today}
\maketitle

\emph{Introduction.}---
Gravitational waves (GWs) produced during inflation provide a powerful probe of the early universe.
For example, GWs can originate from quantum fluctuations of the tensor field, which are directly related to the energy scale of inflation~\cite{Grishchuk:1974ny,Starobinsky:1979ty,Rubakov:1982df,Fabbri:1983us,Abbott:1984fp}.
GWs can also be produced through the bubble collisions associated with a first-order phase transition during inflation~\cite{Jiang:2015qor,Wang:2018caj,An:2020fff,An:2022cce}.
Also, gauge fields could be large sources of the GWs depending on their couplings to the inflaton or some other scalar field~\cite{Barnaby:2010vf,Sorbo:2011rz,Cook:2011hg,Barnaby:2011qe,Anber:2012du,Barnaby:2012xt,Maleknejad:2012fw,Kawasaki:2019hvt,Ozsoy:2020ccy}.
Apart from these sources, scalar perturbations can also induce GWs during inflation.

In most cases, GWs induced by the scalar perturbations during inflation are small because the sources for GW production are at second order in perturbations. 
However, the induced GWs can be large if the sound speed of some scalar field is small~\cite{Biagetti:2013kwa,Biagetti:2014asa,Fujita:2014oba} or scalar perturbations get amplified through some mechanism. 
Previous works have discussed the large induced GWs from scalar field fluctuations resonantly amplified by an oscillating sound speed~\cite{Cai:2018tuh,Cai:2019jah}, oscillatory features in the potential~\cite{Zhou:2020kkf,Cai:2021yvq,Peng:2021zon} or in the noncanonical term~\cite{Cai:2021wzd}, and a rapid turn of the field trajectory~\cite{Fumagalli:2021mpc}.

\begin{figure} 
\centering \includegraphics[width=\columnwidth]{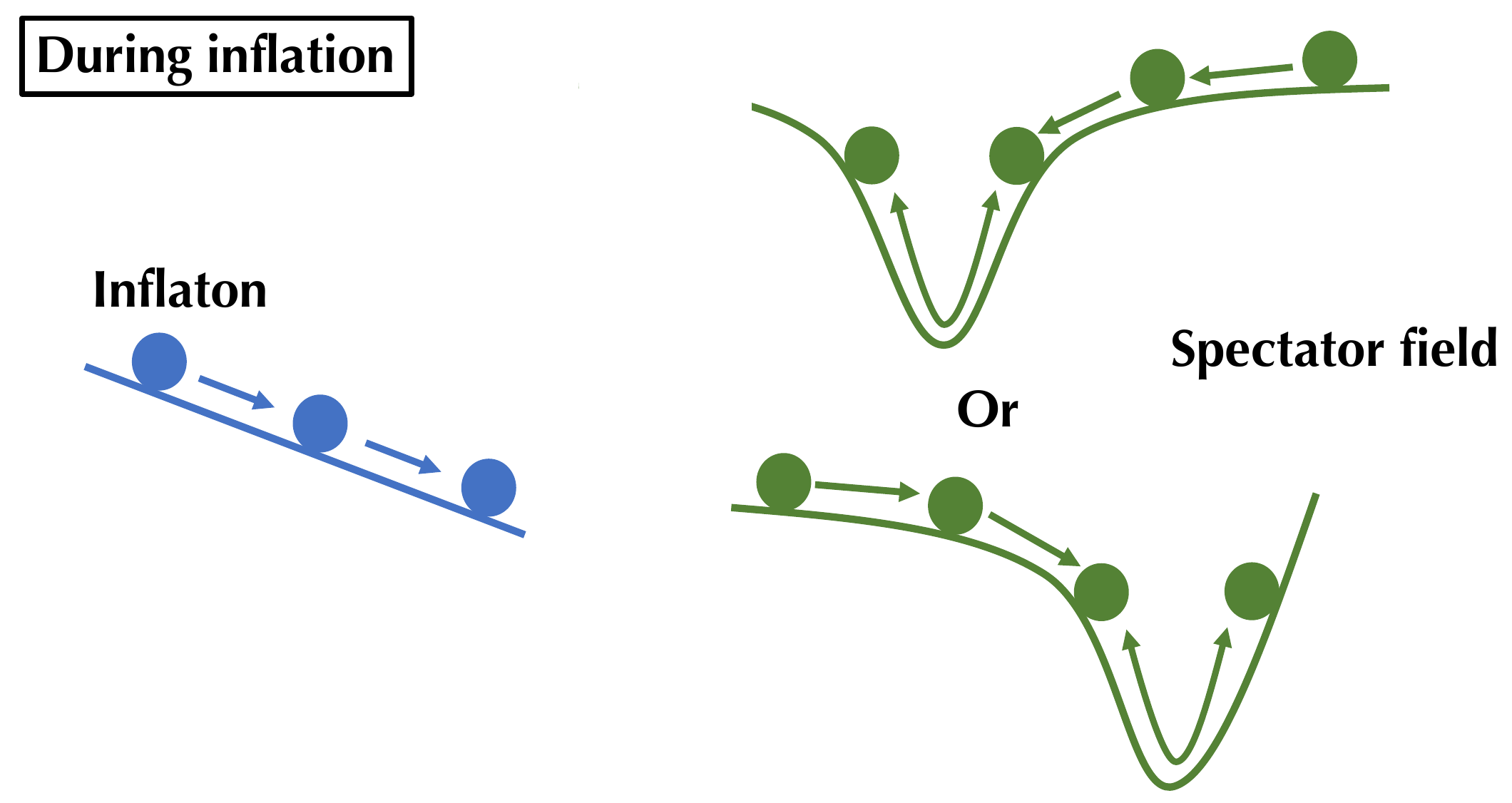}
\caption{ 
The schematic picture of the situation that we focus on throughout this work.
}
\label{fig:pots}
\end{figure}

In this work, we consider the case where a spectator field, not the inflaton, rolls down and oscillates around its potential minimum during inflation.
Figure~\ref{fig:pots} shows the situation we consider.
During the oscillation, the effective mass of the spectator field can also oscillate depending on the potential.
If the oscillation timescale is smaller than the Hubble timescale---that is, the mass of the spectator field around the minimum is heavier than the Hubble parameter---the oscillation can cause the parametric resonance, which amplifies the spectator field fluctuations significantly~\cite{Traschen:1990sw,Dolgov:1989us,Kofman:1994rk,Allahverdi:2010xz}.
Although this amplification phenomenon is often called preheating, we consider the case where the amplification occurs during inflation.
We show that the amplified fluctuations induce large GWs, which could be investigated by future GW observations.
The GW production with the preheating, which occurs after the inflation, has been studied in Refs.~\cite{Khlebnikov:1997di,Garcia-Bellido:1998qth,Easther:2006gt,Easther:2006vd,Garcia-Bellido:2007nns,Garcia-Bellido:2007fiu}.\footnote{Even without the parametric resonance, large GWs could be induced after inflation by some fields other than the inflaton in some cases~\cite{Bartolo:2007vp}.}
On the other hand, in our case, the GWs are induced on subhorizon scales and exit the horizon during inflation.
Finally, after the inflation, the GWs reenter the horizon.
We note that this kind of GW production can occur even if the spectator field has no interaction with any other fields except for gravitational interaction.

\emph{Basic equations.}---
Here, we briefly summarize the basic equations for the scalar field fluctuations and the GWs induced during inflation.
Throughout this work, we take the conformal Newtonian gauge, in which scalar and tensor perturbations are given by 
\begin{align}
  \label{eq:metric_pertb_def}
  \dd s^2 = a^2 \left\{ -(1+2\Phi) \dd \eta^2 + \left[ (1- 2\Phi) \delta_{ij} + \frac{1}{2}h_{ij} \right] \dd x^i \dd x^j \right\},
\end{align}
where $a$ is the scale factor and we have used the perfect fluid condition, which enables us to describe the scalar perturbations with the one parameter, $\Phi$.
For scalar fields, we consider the following action:
\begin{align}
	S = \int \dd^4 x \sqrt{-g} \left[ -\frac{1}{2} \sum_J \partial^\mu \phi_J \partial_\mu \phi_J - V(\phi)  \right],
\end{align}
where $J$ is the index of the field.
From the scalar components of the Einstein equation, we obtain the equation of motion for scalar field fluctuations~\cite{Weinberg:2008zzc}:
\begin{align}
	\delta \phi_J'' + 2\mathcal H \delta \phi_J' - \nabla^2 \delta \phi_J + a^2 \frac{\partial^2 V}{\partial \phi_J \partial\phi_I} \delta \phi^I = -2 a^2 \frac{\partial V}{\partial \phi_J} \Phi + 4 \Phi' \phi_J',
	\label{eq:delta_phi_eom}	
\end{align}
where the prime denotes the derivative with respect to $\eta$ and $\mathcal H \equiv a'/a$. 

From the traceless-transverse component of the Einstein equation, we obtain the equation of motion for tensor perturbations~\cite{Baumann:2007zm,Inomata:2021zel}:
\begin{align}
	{h_{ij}}'' + 2 \mathcal H {h_{ij}}' - \nabla^2 h_{ij} = 4 \hat{\mathcal T}_{ij}^{\ \ lm} S_{lm},
	\label{eq:h_ij_eom}
\end{align}
where the $\hat{\mathcal T}_{ij}$ is the projection operator onto the traceless-transverse space. 
During inflation, the source term, $S_{lm}$, becomes 
\begin{align}
	S_{lm} \simeq& \frac{1}{M_\Pl^2} \sum_J \partial_l \delta \phi_J \partial_m \delta \phi_J \nonumber \\
	\simeq& \frac{1}{M_\Pl^2} \partial_l \delta \chi \partial_m \delta \chi,
\end{align}
where, in the second equality, we have assumed that the fluctuations of the only one field, denoted by $\chi$, get amplified by the parametric resonance for simplicity.

Then, let us go to the Fourier space. 
The tensor perturbations can be expanded with the Fourier modes as 
\begin{align}
	h_{ij}(\bm x) = \sum_{\lambda = +,\times} \int \frac{\dd^3 k}{(2\pi)^3} \ee^{\lambda}_{ij}(\hat k) h^{\lambda}_{\bm k} e^{i \bm k \cdot \bm x}, 
\end{align}
where $\ee^\lambda_{ij}$ is the polarization tensor and $\hat k \equiv \bm k/k$ with $k$ being $|\bm k|$. 
For convenience, we define the transfer function for the Fourier modes of $\delta \chi$ as $\delta \chi_{\bm k}(\eta) = T(k, \eta) \delta \chi_{\bm k}(\eta_*)$, where the transfer function is normalized as $T(k,\eta_*) = 1$ and $\eta_*$ is the conformal time when the power spectrum, $\mathcal P_{\delta \chi}$, at the peak scale becomes maximum due to the parametric resonance.
Note that the power spectrum is related to the ensemble average as 
\begin{align}
	&\expval{\delta \chi_{\bm k}(\eta_*) \delta \chi_{\bm k'}(\eta_*)} = (2\pi)^3 \delta(\bm k + \bm k') \frac{2\pi^2}{k^3} \mathcal P_{\delta \chi}(k,\eta_*).
\end{align}
Using these expressions, we finally obtain the power spectrum of the induced tensor perturbations~\cite{Inomata:2021zel}:
\begin{align}
\label{eq:ph_express_v_u}
\mathcal P_h(k,\eta) =&\frac{4}{M_\Pl^4}
\int^\infty_0 \dd v \int^{|1+v|}_{|1-v|} \dd u \left[ \frac{4v^2 - (1 + v^2 - u^2 )^2}{4uv} \right]^2 \nonumber \\
& \qquad
\times |I(u, v, k, \eta)|^2 \mathcal P_{\delta \chi}(u k, \eta_*) \mathcal P_{\delta \chi} ( v k, \eta_* ),
\end{align}
where $I(u,v,k,\eta)$ is defined as
\begin{align}
  \label{eq:i_vux_def}
  I(u,v,k,\eta) \equiv k^2 \int^\eta_{-\infty} \dd \bar \eta \, g_k(\eta; \bar \eta) T(uk, \bar \eta) T(vk, \bar \eta). 
\end{align}
The $g_k$ is the Green function during inflation, given by 
\begin{align}
	g_k(\eta;\eta') = \Theta(\eta - \eta')& \frac{1}{k^3{\eta'}^2} \left\{ k(\eta' - \eta) \cos[k(\eta' - \eta)] \right.\nonumber \\
	& \left. - (1+ k^2 \eta \eta') \sin[k(\eta' - \eta)] \right\}.
	\label{eq:green}
\end{align}

After the inflation, the induced GWs enter the horizon and contribute to the energy density of the universe. 
Throughout this work, we assume that the induced GWs enter the horizon during the radiation-dominated (RD) era. 
In this case, the current energy density parameter of the induced GWs is given by~\cite{Inomata:2021zel}
\begin{align}
	\Omega_{\GW} (k) h^2 &= 3.4 \times 10^{-7} \left( \frac{g_{*}}{106.75} \right)^{-1/3} \mathcal P_h (k,\eta \rightarrow 0),
\end{align}
where $\Omega_{r,0} h^2 (\simeq 4.18 \times 10^{-5})$ is the current energy density parameter of radiation and $g_{*}$ is the effective degrees of freedom at the horizon entry of the induced GWs.

\emph{Concrete models.}---
To show the mechanism concretely, we consider the following potential as a fiducial example:
\begin{align}
        V(\phi,\chi) = V_1(\phi) + V_2(\chi).
        \label{eq:pot_real}
\end{align}
For the inflaton potential, we consider the following potential,
\begin{align}
        V_1(\phi) &= V_0 (1 - \sqrt{2\epsilon_1} \phi/M_\Pl) + V_\text{end}(\phi).
\end{align}
For the spectator field potential, we consider the following two forms as fiducial choices:
\begin{align}
        V_2(\chi) &= \Delta V \tanh^{2} \left( \frac{\chi}{\sqrt{6\alpha}M_\Pl} \right)  + V_0 \sqrt{2\epsilon_2} \chi/M_\Pl,
        \label{eq:alpha_pot}\\
        V_2(\chi) &= \Delta V \left[1 - \Theta(\chi)\left(\frac{\chi}{\chi_0}\right)^3 \right]^2  - V_0 \sqrt{2\epsilon_2} \chi/M_\Pl,
        \label{eq:hill_pot}
\end{align}
where $\Theta(x)$ is the Heaviside step function.
Note that the potential does not have any interaction terms between the inflaton and the spectator field.
We relate $\epsilon_1$, $V_0$ and $\Delta V$ through $\mathcal P_\zeta(k_\text{CMB}) = (V_0 + \Delta V)/(24 \pi^2 \epsilon_1 M_\Pl^4)$ with $\mathcal P_\zeta(k_\text{CMB}) = 2.1\times 10^{-9}$ to match the Planck results on the CMB scales~\cite{Aghanim:2018eyx}.
Also, we assume that the spectator field does not decay to some radiation perturbatively at least until the end of the GW production.
The $V_\text{end}$ describes the end of inflation and we assume that it is negligible long before then.
The form of $V_\text{end}$ is not relevant to the induced GWs because we consider the case where the GW production occurs long before the end of inflation.
In this work, we restrict ourselves to the case of $\Delta V/V_0 < 1$, where the universe is always dominated by the inflaton potential energy until the end of the slow-roll of the inflaton. 
The terms proportional to $\Delta V$ in $V_2$ determine the oscillation of the spectator field and therefore they are the most important parts of the potentials.
We also note that the first term in Eq.~(\ref{eq:alpha_pot}) is the same as the $\alpha$-attractor inflation potential~\cite{Kallosh:2013yoa} and the first term in Eq.~(\ref{eq:hill_pot}) is the hilltop potential that is predicted in the framework of supergravity~\cite{Kumekawa:1994gx,Izawa:1996dv,Izawa:1997df}, though we have introduced the Heaviside function to simply control the evolution of $\chi$ in the regime of $\chi < 0$, which is irrelevant to the parametric resonance.
On the other hand, the second terms in $V_2$ are not directly related to the mechanism, but we introduce them to keep the spectator field fluctuations from dominating the large-scale curvature perturbations. 
If the tilt of $V_2$ is very small at the horizon exit of some large scales, the spectator field fluctuations can modify the large-scale curvature perturbations after it rolls down to the potential minimum~\cite{Mazumdar:2012rs,Wang:2013oea}.
In this work, we focus on the case where the inflaton fluctuations mainly contribute to the large-scale curvature perturbations for simplicity.

\begin{figure} 
\centering \includegraphics[width=\columnwidth]{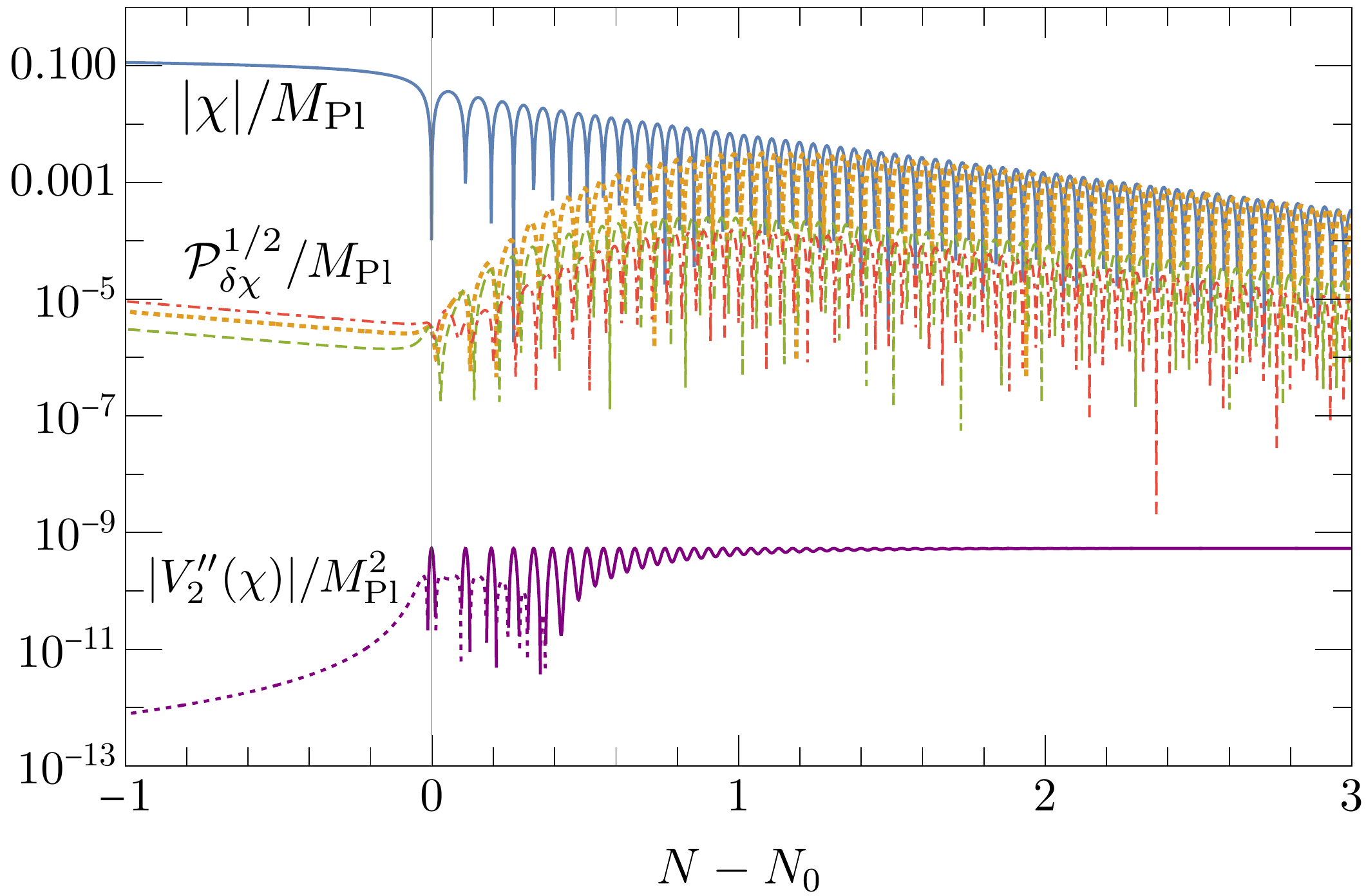}
\caption{ The evolution of $|\chi|/M_\Pl$ (top), $\mathcal P^{1/2}_{\delta\chi}/M_\Pl$ (middle three), and $|V_2''(\chi)|/M_\Pl^2$ (bottom) in the $\alpha$-attractor type potential, given by Eq.~(\ref{eq:alpha_pot}).
We take $\epsilon_1 = 10^{-6}$, $\epsilon_2 = 10^{-4}$, $\alpha = 1.16\times 10^{-4}$, $V_0/M_\Pl^4 = 3.11\times 10^{-13}$, and $\Delta V/V_0=0.6$ for all lines.
For $\mathcal P^{1/2}_{\delta\chi}/M_\Pl$, we take $k = k_{\text{peak},\delta \chi}$ for the orange dotted line, $0.5 k_{\text{peak},\delta \chi}$ for the green dashed line, and $1.5 k_{\text{peak},\delta \chi}$ for the red dot-dashed line, where $k_{\text{peak},\delta \chi} (\simeq  36/|\eta_0|)$ is the peak scale of the resonant amplification of $\delta \chi$.
For $V_2''/M_\Pl^2$, the solid and dotted lines mean $V_2'' > 0$ and $V_2''<0$, respectively.
}
\label{fig:evol}
\end{figure}

\begin{figure} 
\centering \includegraphics[width=\columnwidth]{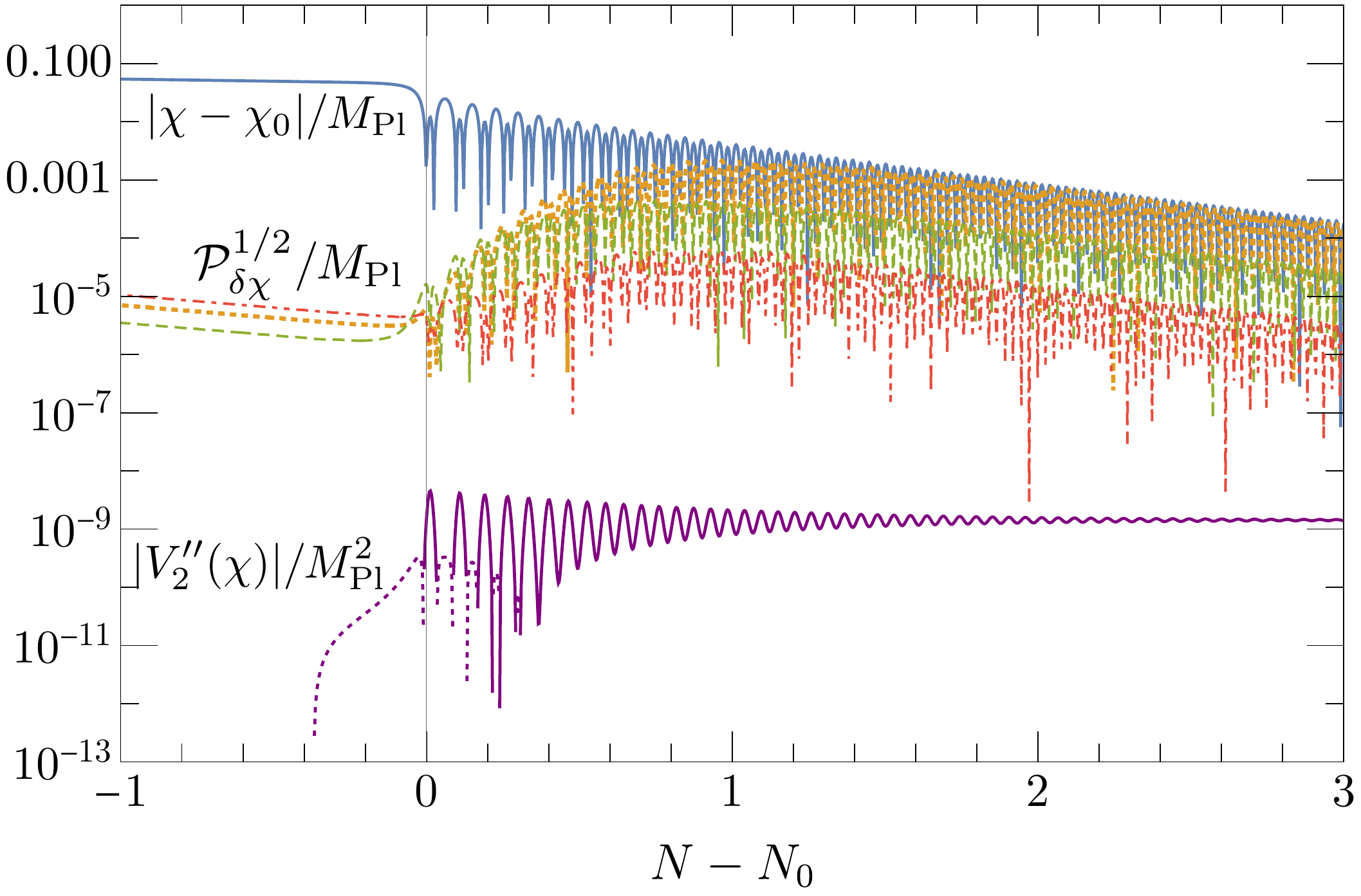}
\caption{ The evolution of $|\chi-\chi_0|/M_\Pl$ (top), $\mathcal P^{1/2}_{\delta\chi}/M_\Pl$ (middle three), and $|V_2''(\chi)|/M_\Pl^2$ (bottom) in the hilltop type potential, given by Eq.~(\ref{eq:hill_pot}).
We take $\chi_0/M_\Pl = 4.85\times 10^{-2}$ and the same values of the other parameters as in Fig.~\ref{fig:evol}.
The peak scale is given by $k_{\text{peak},\delta \chi} \simeq  41/|\eta_0|$.
}
\label{fig:evol_hill}
\end{figure}

Figures~\ref{fig:evol} and ~\ref{fig:evol_hill} show the evolution of the background, the perturbations, and the effective mass of the spectator field in the two types of the spectator field potential, Eqs.~(\ref{eq:alpha_pot}) and (\ref{eq:hill_pot}), respectively. 
The spectator field starts to oscillate around its potential minimum, $\chi \simeq 0$ or $\chi \simeq \chi_0$, at some point. 
We define the first e-folds and the conformal time at $\chi = 0$ or $\chi = \chi_0$ as $N_0$ and $\eta_0$, respectively.
During the oscillation, the effective mass of the spectator field, $V_2''$, also oscillates, which causes the parametric resonance for the spectator field fluctuations whose wavenumber is close to the timescale of the oscillation.
The oscillation amplitude of the spectator field decreases proportionally to $a^{-3/2}$ due to the expansion of the universe.
Accordingly, the oscillation amplitude of the effective mass also decreases, which shuts off the resonance along with the expansion of the physical wavelength of the peak-scale field fluctuations.
After the resonance, the spectator field fluctuations also decrease as $\mathcal P_{\delta \chi}^{1/2} \propto a^{-3/2}$. 
From this behavior of $\mathcal P_{\delta \chi}$, we can see that the GWs are mainly induced around the end of the resonance. 
In the fiducial parameter sets, the oscillation amplitude of $\mathcal P^{1/2}_{\delta \chi}(k_{\text{peak},\delta \chi})$, which corresponds to the typical amplitude of $\delta \chi$, is smaller than that of $|\chi|$ or $|\chi-\chi_0|$. 
Strictly speaking, when $\mathcal P^{1/2}_{\delta \chi}(k_{\text{peak},\delta \chi})$ is comparable to $|\chi|$ or $|\chi-\chi_0|$, the backreaction, which we do not take into account, could affect their evolution.
However, the resonant amplification mainly occurs during $\mathcal P^{1/2}_{\delta \chi}(k_{\text{peak},\delta \chi}) \ll |\chi|$ or $|\chi-\chi_0|$, and therefore we can expect that the order of the amplification would not change even if we take into account the backreaction.

\begin{figure} 
\centering \includegraphics[width=\columnwidth]{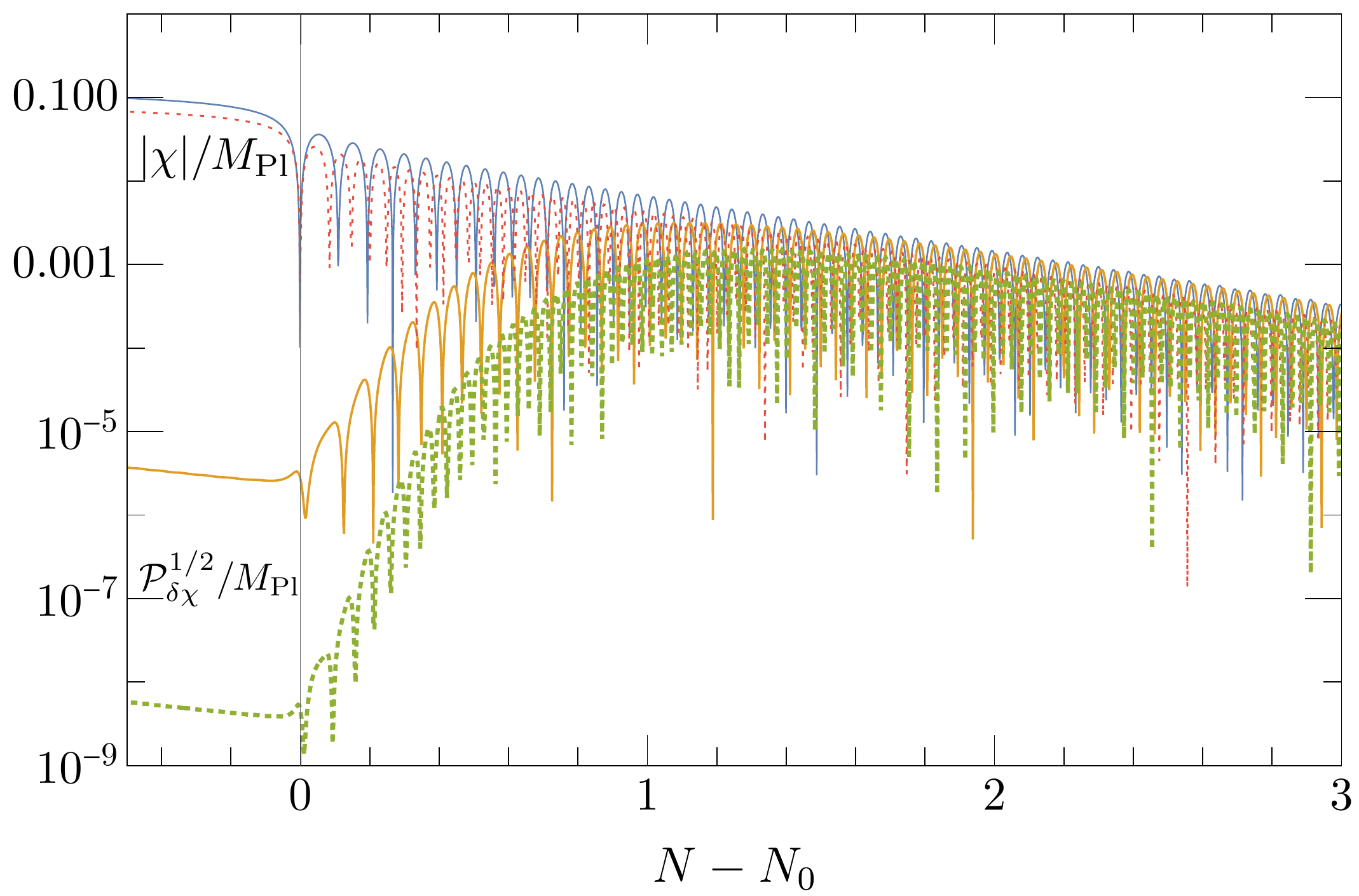}
\caption{ 
The comparison between the cases of $V_0/M_\Pl^4 = 3.11\times 10^{-13}$ (solid lines) and $V_0/M_\Pl^4 = 3.11\times 10^{-19}$ (dotted lines) in the $\alpha$-attractor potential.
The upper lines show the background evolution of the spectator field and the lower lines show $\mathcal P^{1/2}_{\delta\chi}(k_{\text{peak},\delta \chi})/M_\Pl$.
For the case of $V_0/M_\Pl^4 = 3.11\times 10^{-13}$ (solid lines), we take the same parameters as in Fig.~\ref{fig:evol}.
For the case of $V_0/M_\Pl^4 = 3.11\times 10^{-19}$ (dotted lines), we take the other parameters as $\epsilon_1 = 10^{-12}$, $\epsilon_2 = 10^{-10}$, $\alpha = 4.4\times 10^{-5}$, $\Delta V/V_0=0.6$, and $k_{\text{peak},\delta \chi} \simeq  55/|\eta_0|$.
}
\label{fig:evol_ep_comp}
\end{figure}

As we will see, the strength of the parametric resonance is sensitive to $\alpha$ or $\chi_0$, which determines the ratio between the spectator field mass and the Hubble parameter if $\Delta V/V_0$ is fixed. 
Specifically, the ratio can be expressed as $m_\chi/H|_{\eta=\eta_0} \simeq \sqrt{\frac{\Delta V}{\alpha(V_0+\Delta V)}}$ in the $\alpha$-attractor potential (Eq.~(\ref{eq:alpha_pot})) and $m_\chi/H|_{\eta=\eta_0} \simeq 3\sqrt{6} \frac{M_\Pl}{\chi_0} \sqrt{\frac{\Delta V}{V_0+\Delta V}}$ in the hilltop potential (Eq.~(\ref{eq:hill_pot})), where $m_\chi$ is the spectator field mass around the potential minimum ($\chi =0$ or $\chi = \chi_0$).\footnote{If $\alpha$ or $\chi_0$ is fixed, the $m_\chi/H$ is determined by $\Delta V/V_0$ and therefore the resonance amplification sensitively depends on $\Delta V/V_0$, instead.}
The sensitive dependence leads to the fact that, even if we consider a smaller energy scale of inflation, the amplified fluctuations can still be large with a smaller $\alpha$ or $\chi_0$.
Here, let us see this behavior in the case of the $\alpha$-attractor potential.
The dotted lines in Fig.~\ref{fig:evol_ep_comp} show the evolution of the quantities in the $\alpha$-attractor potential with a smaller $V_0 (=3.11 \times 10^{-19})$. 
For the dotted lines, we take a bit smaller value of $\alpha$ to make the $\delta \chi$ close to the backreaction upper bound, $\mathcal P_{\delta \chi}^{1/2} \sim |\chi|$, similar to the case of Fig.~\ref{fig:evol}.
From Fig.~\ref{fig:evol_ep_comp}, we can see that the amplified fluctuations can be of the same order of magnitude even in the small inflation energy scale.

\begin{figure} 
\centering \includegraphics[width=\columnwidth]{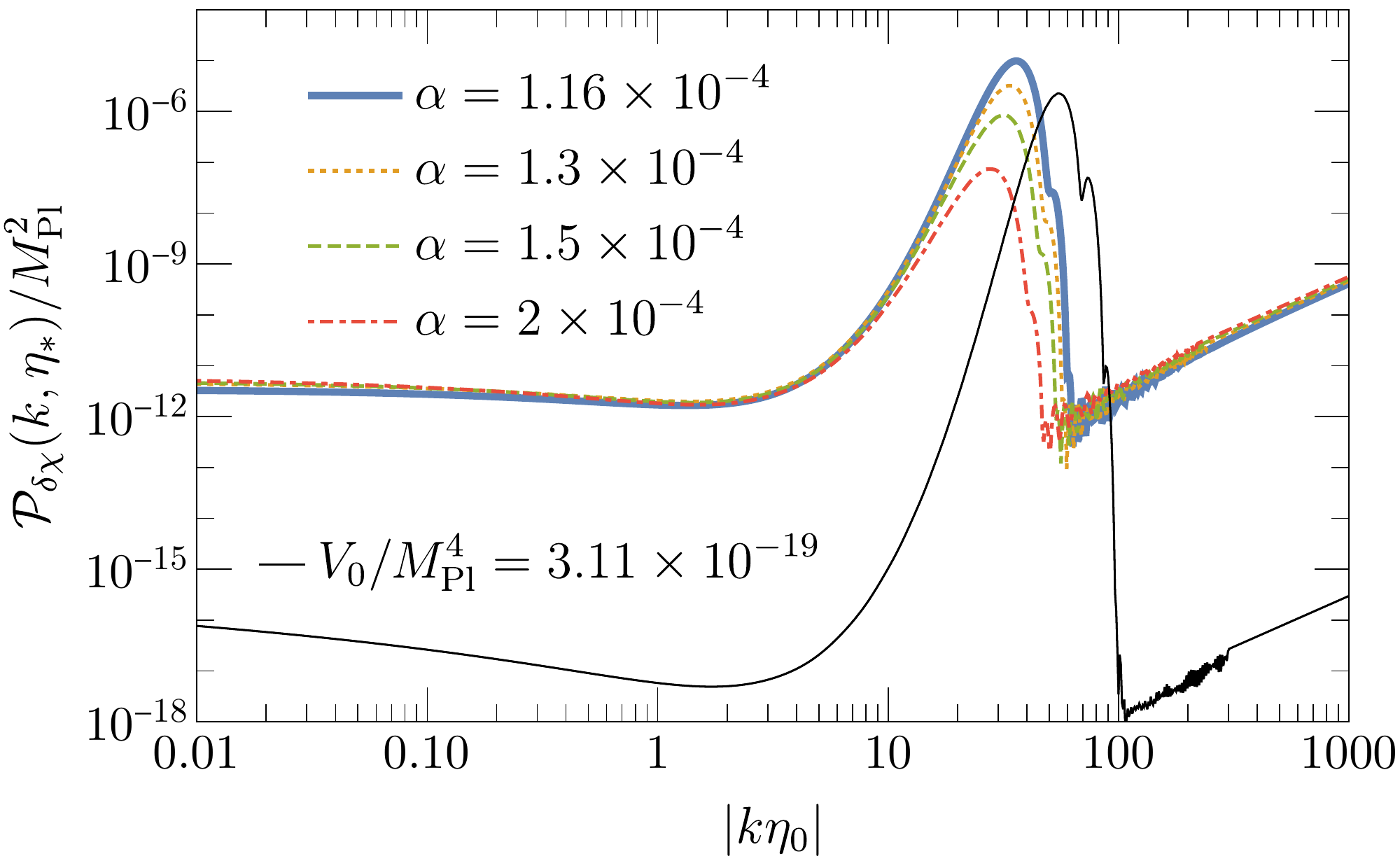}
\caption{ 
The power spectrum of the spectator field fluctuations in the $\alpha$-attractor potential at the time when $\mathcal P_{\delta \chi}(k_{\text{peak},\delta \chi})$ reaches its maximum, denoted by $\eta_*$.
For blue solid line, we take the same parameters as in Fig.~\ref{fig:evol}. 
For the dotted, dashed, dot-dashed lines, we take the different values of $\alpha$, but the same values for the other parameters as in Fig.~\ref{fig:evol}.
For the black thin line, we take the same parameters as the dotted lines in Fig.~\ref{fig:evol_ep_comp}.
The e-folds at $\eta_*$, denoted by $N_*$, become $N_*-N_0 = 1.12$ for $\alpha = 1.16\times 10^{-4}$, $N_*-N_0 = 1.08$ for $\alpha = 1.3\times 10^{-4}$, $N_*-N_0 = 1.03$ for $\alpha = 1.5\times 10^{-4}$, and $N_*-N_0 = 0.969$ for $\alpha = 2\times 10^{-4}$.
For the case of $V_0/M_\Pl^4 = 3.11\times 10^{-19}$, we have $N_*-N_0 = 1.31$.
}
\label{fig:pchi}
\end{figure}

\begin{figure} 
\centering \includegraphics[width=\columnwidth]{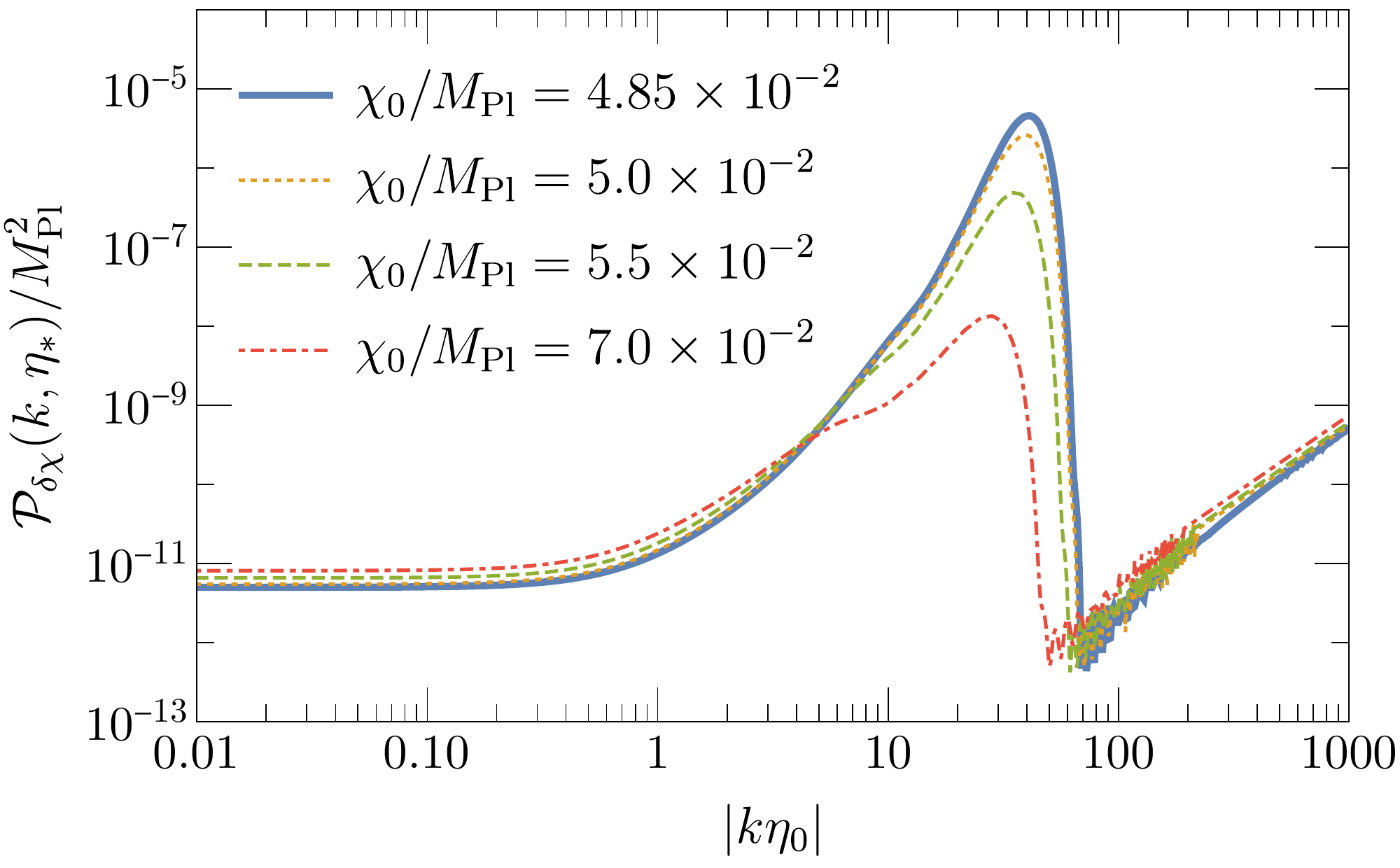}
\caption{ 
The power spectrum of the spectator field fluctuations when $\mathcal P_{\delta \chi}(k_{\text{peak},\delta \chi})$ reaches its maximum in the hilltop potential.
We take different values of $\chi_0$ and the same values for the other parameters as in Fig.~\ref{fig:evol_hill}.
We have $N_*-N_0 = 1.02$ for $\chi_0/M_\Pl = 4.85\times 10^{-2}$, $N_*-N_0 = 0.991$ for $\chi_0/M_\Pl = 5.0\times 10^{-2}$, $N_*-N_0 = 0.948$ for $\chi_0/M_\Pl = 5.5\times 10^{-2}$, and $N_*-N_0 = 0.839$ for $\chi_0/M_\Pl = 7.0\times 10^{-2}$.
}
\label{fig:pchi_hill}
\end{figure}

Figures~\ref{fig:pchi} and \ref{fig:pchi_hill} show the power spectrum of the spectator field fluctuations at the time when the peak-scale power spectrum reaches its maximum in the $\alpha$-attractor and the hilltop potential, respectively.
From these figures, we can see that the peak value becomes $\mathcal O(10^{-5})$ at that time for the parameter sets in Figs.~\ref{fig:evol} and \ref{fig:evol_hill}. 
The increase on the very small scales ($|k\eta_0| \gtrsim 100$) is due to the ordinary subhorizon evolution that is not affected by the parametric resonance, which leads to $\mathcal P_{\delta \chi} \propto (k/a)^{2}$.
For comparison, we also show the results with larger values of $\alpha$ or $\chi_0$, which correspond to longer oscillation timescales. 
In this case, the peak height becomes smaller because the resonance ends after fewer oscillations.
Also, by comparing the fiducial result with $V_0/M_\Pl^4 = 3.11\times 10^{-13}$ (blue line) and the result with $V_0/M_\Pl^4 = 3.11\times 10^{-19}$ (black thin line) in Fig.~\ref{fig:pchi}, we can see again that the amplified fluctuations can be of the same order of magnitude even in the smaller inflation energy scale.

\begin{figure} 
\centering \includegraphics[width=\columnwidth]{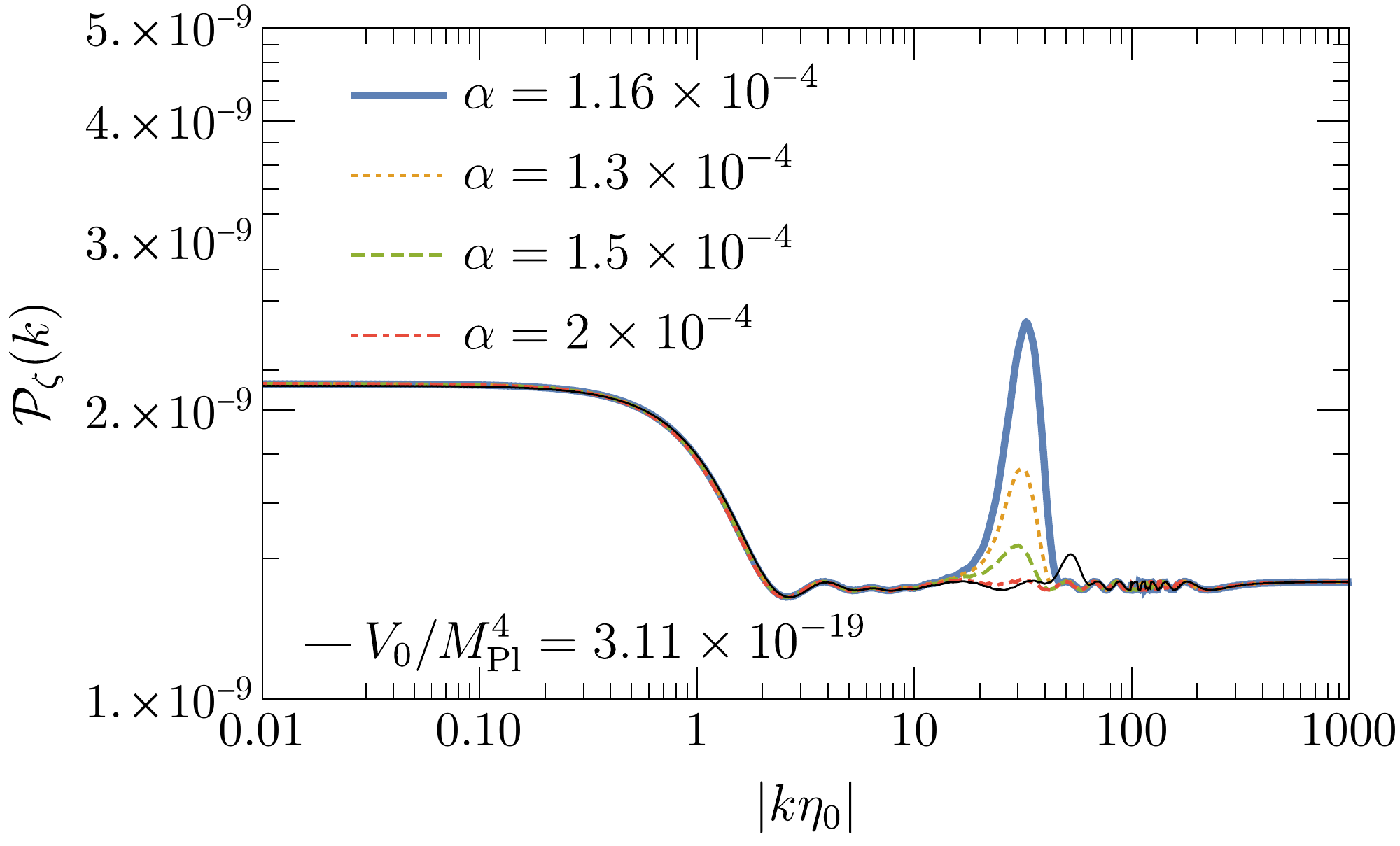}
\caption{ 
The final power spectrum of curvature perturbations after the spectator field fluctuations become negligible due to their damping during its oscillation in the $\alpha$-attractor potential.
The parameters are the same as in Fig.~\ref{fig:pchi}.
}
\label{fig:pzeta}
\end{figure}

\begin{figure} 
\centering \includegraphics[width=\columnwidth]{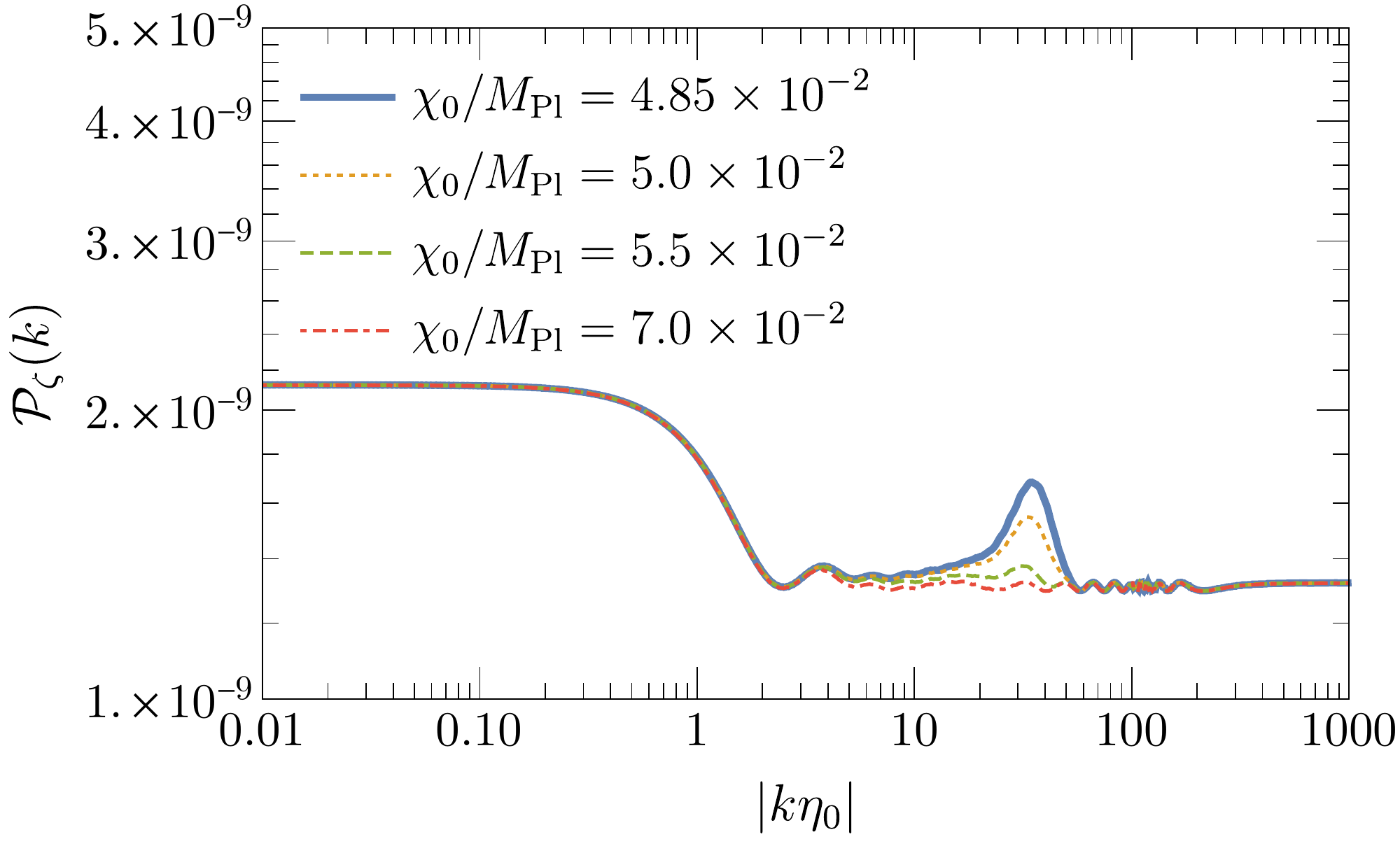}
\caption{ 
The final power spectrum of curvature perturbations in the hilltop potential.
The parameters are the same as in Fig.~\ref{fig:pchi_hill}.
}
\label{fig:pzeta_hill}
\end{figure}

Figures~\ref{fig:pzeta} and \ref{fig:pzeta_hill} show the power spectrum of curvature perturbations with the same parameters as in Figs.~\ref{fig:pchi} and ~\ref{fig:pchi_hill}.
Although, as shown in Figs.~\ref{fig:evol} and \ref{fig:evol_hill}, the amplified spectator field fluctuations damp away after a while, the inflaton fluctuations partially inherit the amplified fluctuations through gravitational interaction. 
Specifically, the inflaton fluctuations are affected through the gravitational potential $\Phi$ in Eq.~(\ref{eq:delta_phi_eom}).
This is why the power spectrum exhibits the small peak around $|k\eta_0| \simeq 30$. 
Also, the decay of the normalization around $|k\eta_0|=1$ is due to the change of the total potential energy as $V_0+\Delta V \rightarrow V_0$.
If the curvature perturbations enter the horizon during a RD era, the GWs induced by the curvature perturbations around the horizon entry are $\Omega_\GW h^2 \simeq \mathcal O(10^{-5} \mathcal P_\zeta^2) \simeq \mathcal O(10^{-23})$ (see e.g. Refs.~\cite{Espinosa:2018eve,Kohri:2018awv,Domenech:2021ztg} for the GWs induced during the RD era).
This is much smaller than those induced during the inflation, which are shown in the next figures.

\begin{figure} 
\centering \includegraphics[width=\columnwidth]{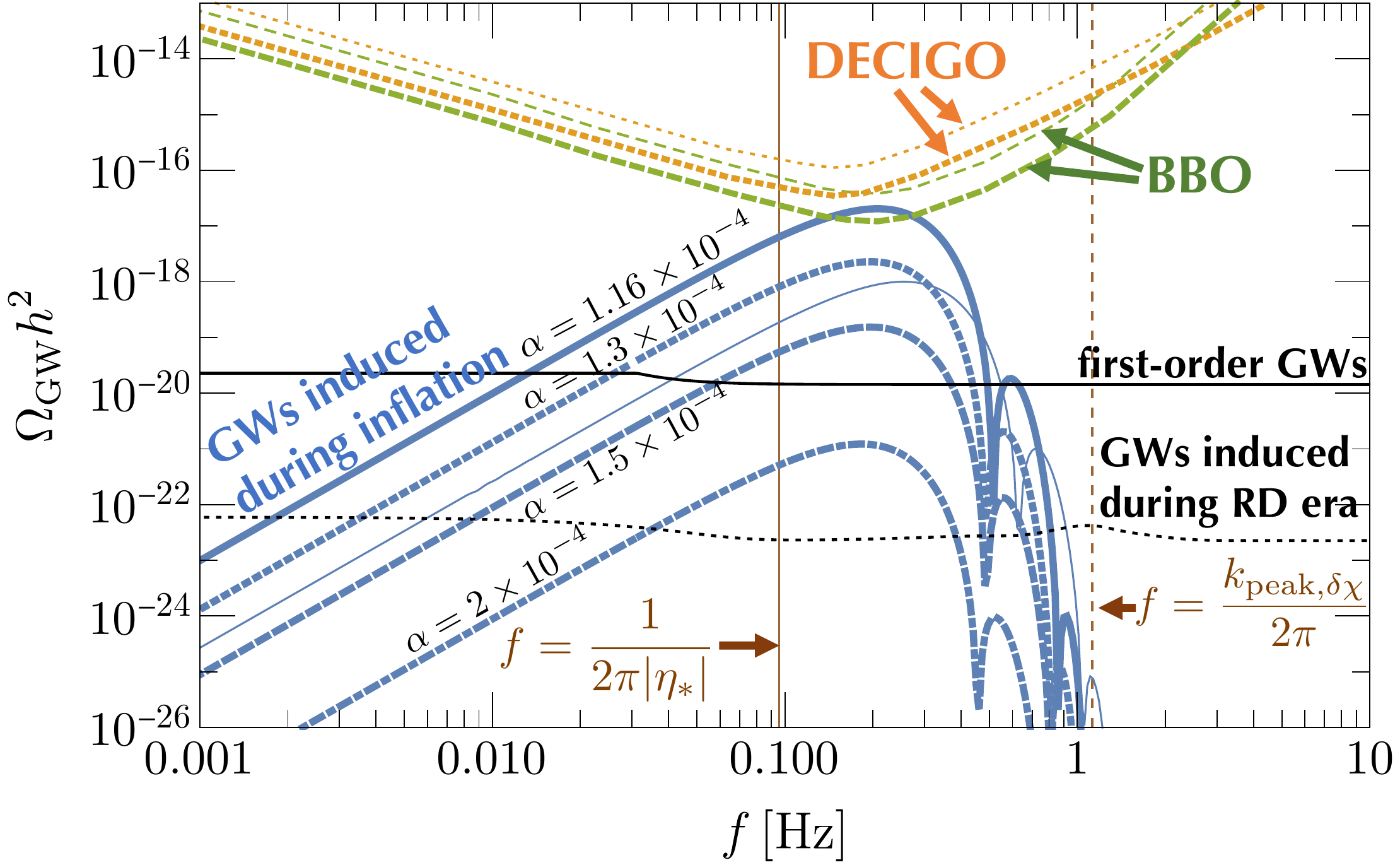}
\caption{ 
The spectrum of induced GWs in the $\alpha$-attractor potential (blue lines).
We take $|\eta_0|^{-1} = 2.0\times 10^{13}$\,Mpc$^{-1}$ and the parameters in Fig.~\ref{fig:pchi}.
The thin blue line (between the lines for $\alpha=1.3\times 10^{-4}$ and $1.5\times 10^{-4}$) corresponds to the case of $V_0/M_\Pl^4 = 3.11 \times 10^{-19}$.
The black thin solid line shows the first-order GWs in the case of $V_0/M_\Pl^4 = 3.11 \times 10^{-13}$.
The black dashed line shows the GWs induced by the curvature perturbations during the RD era.
The two brown solid and dashed vertical lines show the frequencies that correspond to the horizon scale at $\eta_*$ and the peak scale of $\delta \chi$ in the case of $\alpha = 1.16\times 10^{-4}$, respectively.
For comparison, we also plot the effective sensitivity curves for DECIGO (dotted line) and BBO (dashed line) and the thin curves are for $1$-year observation and the thick ones for $10$-year observation.
These curves are based on Refs.~\cite{Thrane:2013oya,Kuroyanagi:2014qza,Inomata:2018epa} and, in particular, the thin sensitivity curves are the same as the ones in Fig.~1 of Ref.~\cite{Inomata:2018epa}. 
}
\label{fig:gw}
\end{figure}

\begin{figure} 
\centering \includegraphics[width=\columnwidth]{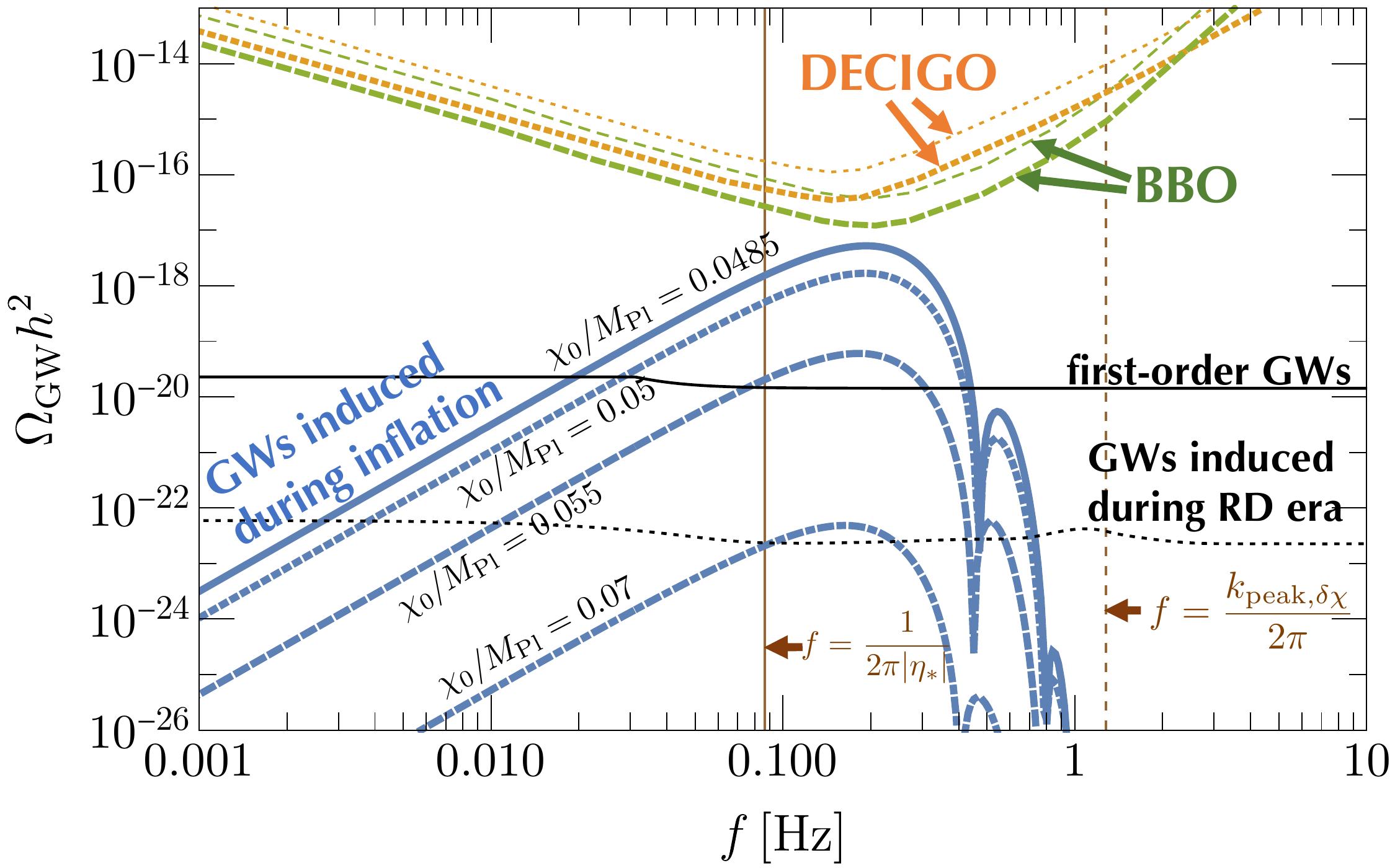}
\caption{ 
The spectrum of induced GWs in the hilltop potential (blue lines).
We take $|\eta_0|^{-1} = 2.0\times 10^{13}$\,Mpc$^{-1}$ and the parameters in Fig.~\ref{fig:pchi_hill}.
The two brown solid and dashed vertical lines show the frequencies in the case of $\chi_0/M_\Pl = 0.0485$.
}
\label{fig:gw_hill}
\end{figure}

Figures~\ref{fig:gw} and \ref{fig:gw_hill} show the GW spectrum with $|\eta_0|^{-1} = 2.0\times 10^{13}$\,Mpc$^{-1}$ and the parameters taken in Figs.~\ref{fig:pchi} and \ref{fig:pchi_hill}.
To reduce the computational cost of the integrals in Eq.~(\ref{eq:ph_express_v_u}), we have approximated $T(k,\eta) \simeq T(k_{\text{peak},\delta \chi},\eta)$ given the fact that the dominant contribution comes from the scales around $k_{\text{peak},\delta \chi}$ and the phases of the oscillations around the peak scale are almost the same (see Figs.~\ref{fig:evol} and \ref{fig:evol_hill}). 
At the peak scale in the smallest $\alpha$ and $\chi_0$ cases, $\Omega_\text{GW}(k_{\text{peak,GW}}) h^2 \simeq \mathcal O(10^{-17})$, which corresponds to $\mathcal P_h(k_{\text{peak,GW}},\eta \rightarrow 0) \simeq \mathcal O(10^{-11})$.
Note that the peak value of $\mathcal P_h$ is of $\mathcal O(\mathcal P^2_{\delta \chi} (k_{\text{peak},\delta \chi}, \eta_*))$.
The peak scale of GWs, $k_\text{peak,GW}$, is close to the horizon scale at $\eta_*$ and generally different from the peak scale of the resonant amplification of $\delta \chi$, $k_{\text{peak},\delta \chi}$.
This behavior is consistent with the results in Refs.~\cite{Cai:2019jah,Inomata:2021zel}.
For comparison, we also show the first-order GWs, whose power spectrum is given by $\mathcal P_h = 4H^2/(\pi M_\Pl)^2$, and the GWs induced by the curvature perturbations during the RD era.
We can see that the GWs induced during inflation can be larger than the other GWs around the peak scale.
These figures also indicate that the induced GWs could possibly be investigated by future projects, such as DECIGO and BBO.
(See also Ref.~\cite{Sedda:2019uro} for the summary of the GW projects around $f\sim \mathcal O(0.1)$\,Hz.)

\emph{Necessary conditions.}---
Here, let us discuss the necessary conditions for the sufficient enhancement of the induced GWs with the fast oscillation of the spectator field during inflation.

First, the oscillation timescale must be smaller than the Hubble timescale.  
Otherwise, the damping of the oscillation amplitude of the spectator field shuts off the resonance before it amplifies the fluctuations sufficiently.

Second, the oscillation timescale must not be too small.
This condition is related to the energy conservation law.
The energy density of the amplified $\delta \chi$ is given by $\rho_{\delta \chi} \simeq \expval{(\delta \chi')^2 + (\partial_i \delta \chi)^2 + a^2 V''_2(\chi) \delta \chi^2}/(2a^2)$~\cite{Inomata:2021zel}.
Here, for simplicity, let us define $m^2_\chi$ as the value of $V_2''(\chi)$ around the minimum and assume that the effective mass oscillation occurs with the timescale of $\mathcal O(1/m_\chi)$.
Then, given that the resonance peak scale is $(k_{\text{peak},\delta \chi}/a(\eta_0))^2 \simeq \mathcal O(m_\chi^2)$, the energy density can be naively approximated as $\rho_{\delta \chi} \simeq \mathcal O(m_\chi^2\expval{\delta \chi^2})$.
Since the origin of $\rho_{\delta \chi}$ is the potential energy of the spectator field $\Delta V$, the $\rho_{\delta \chi}$ is upper bounded as $\rho_{\delta \chi} < \Delta V$.
Given $\Delta V$, a smaller $m_\chi$ gives a larger upper bound on $\delta \chi$, which allows larger induced GWs~\cite{Inomata:2021zel}.
For this reason, $m_\chi$ must not be too large compared to the Hubble parameter.
For example, our fiducial parameter sets in Figs.~\ref{fig:evol} and \ref{fig:evol_hill} give $m_\chi/H|_{\eta=\eta_0} \simeq \sqrt{\frac{\Delta V}{\alpha(V_0+\Delta V)}}= 57$ in the $\alpha$-attractor potential and $m_\chi/H|_{\eta=\eta_0} \simeq 3\sqrt{6} \frac{M_\Pl}{\chi_0} \sqrt{\frac{\Delta V}{V_0+\Delta V}} = 93$ in the hilltop potential, respectively.
Note that the mass in the hilltop potential is almost twice as large as that in the $\alpha$-attractor because the $\alpha$-attractor potential leads to two oscillations of the mass during one oscillation of the background field, while the hilltop potential leads to only one mass oscillation during one background oscillation.
Although there is still the possibility that a larger value of $m_\chi/H$ maximizes the induced GWs, the determination of the value requires the lattice simulation because the backreaction and the non-perturbative effects would be important for the larger $m_\chi/H$, which is beyond the scope of this work.

Third, related to the second condition, $\Delta V/V_0$ must be large enough.
The slow-roll parameter at $N = N_0$ becomes $\epsilon_0 \equiv -(\mathcal H' - \mathcal H^2)/\mathcal H^2|_{\eta=\eta_0} \simeq 3\Delta V/V_0$ in the case of $\Delta V/V_0 \ll 1$. 
According to the results in Ref.~\cite{Inomata:2021zel}, the upper bound on the induced GWs is $\mathcal P_h(k_\text{peak,GW}) < \mathcal O(\epsilon_0^2 /(k_{\text{peak},\delta \chi} \eta_0)^2)$.
The reason why the upper bound depends on the ratio $\Delta V/V_0$ (that is, $\epsilon_0$), not on $\Delta V$ itself, is that the amplitude of $\delta \chi$ cannot overcome the initial amplitude of the background oscillation due to the energy conservation.
The initial oscillation amplitude is mainly determined by $\epsilon_0$ if we fix the ratio between the mass and the Hubble parameter.
Because of this, the value of $\Delta V/V_0$ needs to be large for large GW production.
However, if $\Delta V/V_0 > 1$, the universe expansion is temporarily dominated by the spectator field oscillation for a while, which leads to a matter-dominated era sandwiched between two inflation periods.
In this case, the Green function, Eq.~(\ref{eq:green}), must be modified. 
We leave the calculation of the induced GWs in this case for future work. 

Fourth, the potential of the spectator field must lead to the oscillation of the spectator field's effective mass.
To this end, the fast oscillation, whose timescale is smaller than the Hubble timescale, must start before the evolution of the spectator field completely follows the quadratic potential around the minimum, $\sim m_\chi^2 \chi^2/2$. 
The deviation from the quadratic potential leads to the mass oscillation, though the spectator field evolution finally follows the quadratic potential around the minimum  after a while in most cases.
Our fiducial potentials automatically satisfy this condition if the mass around the minimum is heavy enough.
Note that this necessary condition could be relaxed if the spectator field has some interaction with another field. 
For example, if the spectator field has an interaction given by $\lambda \chi^2 \varphi^2$, the oscillation of $\chi$ leads to the mass oscillation of $\varphi$, which can cause the parametric resonance of $\delta \varphi$ in principle. 
Even in this case, large GWs could be induced.
We leave the detailed analysis of this possibility for future work.

Finally, let us mention the relation between these conditions and the parameters in our setups. 
In our fiducial setups, the conditions are directly related to the parameters, $\Delta V/V_0$ and $\alpha$ in the $\alpha$-attractor potential and $\chi_0$ in the hilltop potential, while the conditions are irrelevant to $\epsilon_2$ because the $\epsilon_2$ term is just introduced to suppress the spectator field contributions to the CMB-scale power spectrum.
On the other hand, $\epsilon_1$ is indirectly related to the second and third conditions. 
Since the inflation energy scale itself determines the initial amplitude of the spectator field fluctuations at the beginning of the resonance, a smaller $\epsilon_1$, that is a smaller $V_0$, requires a stronger resonance with a larger $m_\chi/H$ for the large GW production.
From the second and the third conditions, this situation in a smaller $V_0$ (or $\epsilon_1$) leads to a stronger upper bound on the GWs because of a larger $k_{\text{peak},\delta \chi}$.
We can see this situation even if we restrict ourselves to the linear regime.
In Figs.~\ref{fig:evol_ep_comp} and \ref{fig:pchi}, we can see that, in the smaller $V_0$, the resonance peak becomes slightly smaller and shifts to the smaller scale (the larger $k_{\text{peak},\delta \chi}$) even if we tune $\alpha$ to make the $\delta \chi$ close to the backreaction upper bound, $\mathcal P_{\delta \chi}^{1/2} \sim |\chi|$.
Although the $\epsilon_1$ dependence is weak, we need to be careful about the $\epsilon_1$ dependence of the GW upper bound in the case where $\epsilon_1$ is much smaller than our fiducial values.

\emph{Conclusion.}---
In this work, we have discussed the GWs induced by a heavy spectator field that starts to oscillate during inflation.
In some potentials, the effective mass of the spectator field also oscillates during the oscillation period, which could amplify the field fluctuations on some peculiar scales resonantly.
We have shown that the amplified fluctuations can induce large GWs, which could possibly be investigated by future projects, such as BBO and DECIGO.
This kind of induced GWs could be used as detectable traces of a heavy field that is negligible in energy density at the present and is not coupled to any other fields except through gravitational interaction.

\emph{Acknowledgments.}---
The author thanks Jose Mar\'{i}a Ezquiaga, Wayne Hu, Austin Joyce, Hayden Lee, Misao Sasaki, and Lian-Tao Wang for helpful comments on this work.
The author was supported by the Kavli Institute for Cosmological Physics at the University of Chicago through an endowment from the Kavli Foundation and its founder Fred Kavli.

\small
\bibliographystyle{apsrev4-1}
\bibliography{induced_gw_reso}

\end{document}